\documentclass[
reprint,
superscriptaddress,
showkeys,
 amsmath,amssymb,
 aps,
pra,
floatfix,
]{revtex4-1}

\usepackage{graphicx}
\usepackage{dcolumn}
\usepackage{bm}

\usepackage{siunitx}

\usepackage{xcolor}
\usepackage[normalem]{ulem}

\begin{document}
\setlength{\belowcaptionskip}{-10pt}
\raggedbottom
\parskip=5pt

\title{Azetidinium as Cation in Lead Mixed Halide Perovskite Nanocrystals of Optoelectronic Quality}%

\author{Sameer Vajjala \surname{Kesava}}
\email[Correspondence: ]{sameer.vajjalakesava@physics.ox.ac.uk,}
\affiliation{Department of Physics, University of Oxford, OX1 3PU, England, UK}
\author{Yasser \surname{Hassan}}
\author{Alberto \surname{Privitera}}
\affiliation{Department of Physics, University of Oxford, OX1 3PU, England, UK}
\author{Aakash \surname{Varambhia}}
\affiliation{Department of Materials, University of Oxford, OX2 6HT, England, UK}
\author{Henry J. \surname{Snaith}}
\author{Moritz K. \surname{Riede}}
\email[]{moritz.riede@physics.ox.ac.uk} 
\affiliation{Department of Physics, University of Oxford, OX1 3PU, England, UK}

\date{\today}%

\begin{abstract}
Previous theoretical calculations show azetidinium has the right radial size to form a 3D perovskite with lead halides \cite{Kieslich2014_ChemSci}, and has been shown to impart, as the A-site cation of ABX$_{3}$ unit, beneficial properties to ferroelectric perovskites \cite{Zhou2011_AngChem}. However, there has been very limited research into its use as the cation in lead halide perovskites to date. In this communication we report the synthesis and characterization of azetidinium-based lead mixed halide perovskite colloidal nanocrystals. The mixed halide system is iodine and chlorine unlike other reported nanocrystals in the literature where the halide systems are either iodine/bromine or bromine/chlorine. UV-visible absorbance data, complemented with photoluminescence spectroscopy, reveals an indirect-bandgap of about 1.96 eV for our nanocrystals. Structural characterization using TEM shows two distinct interatomic distances (2.98 $\pm$ 0.15 \si{\angstrom} and 3.43 $\pm$ 0.16 \si{\angstrom}) and non-orthogonal lattice angles ($\approx$ 112\si{\degree}) intrinsic to the nanocrystals with a probable triclinic structure revealed by XRD. The presence of chlorine and iodine within the nanocrystals is confirmed by EDS spectroscopy. Finally, light-induced electron paramagnetic resonance (LEPR) spectroscopy with PCBM confirms the photoinduced charge transfer capabilities of the nanocrystals. The formation of such semiconducting lead mixed halide perovskite using azetidinium as the cation suggests a promising subclass of hybrid perovskites holding potential for optoelectronic applications such as in solar cells and photodetectors.  
\end{abstract}

\maketitle

\noindent
\textbf{Introduction.} Organic/inorganic lead halide perovskites (LHP), using earth-abundant elements, have shown tremendous potential in achieving lab-scale efficiencies of solar cells approaching that of the crystalline silicon solar cells \cite{NREL_2018}. The current high-performing LHPs are composed of methylammonium (MA), formamidinium (FA) and caesium (Cs) as the A-site cations in ABX$_{3}$, which forms a 3D structure. This dimensionality, i.e. 3D, 2D or 1D, estimated in terms of the Goldschmidt Tolerance Factor, is dependent on the size of the cation with all others remaining the same.\hspace*{\fill}

\begin{figure}[t]
\includegraphics[scale=0.45]{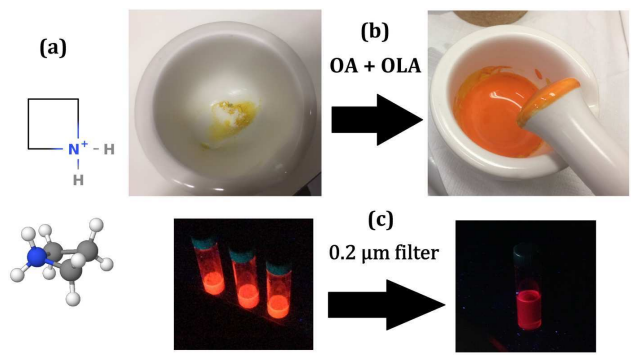}
\caption{\label{fig:Optical}(a). Molecular structure of azetidinium in 2D and 3D (drawn in MolView \cite{MolView}). The 3D structure highlights the restricted rotational motion of the molecule unlike with MA and FA which possess 3D rotational degrees of freedom due to the unrestricted motion of the atoms. (b). Synthesis of AzPbI$_{2}$Cl: addition of oleic acid (OA) and oleyl amine (OLA) to a mixture of PbI$_{2}$ (yellow) and (CH$_{2}$)$_{3}$NH$_{2}$Cl (white), and mixing results in the formation of an orange-colored compound. (c) left: the resultant product (CS) in toluene; and right: filtrate from CS filtered  with a 0.2 $\mu$m PTFE filter, both luminescing in a UV-box (365 nm).} 
\end{figure}

\noindent
In this context, the extended Tolerance Factor approach developed by Kieslich \textit{et al.} has shown azetidinium (Az; Figure \ref{fig:Optical}(a)) as a prospective A-site cation with a Tolerance Factor of 0.98 with lead iodide and radial size MA $<$ Az $<$ FA \cite{Kieslich2014_ChemSci, Kieslich2015_ChemSci}. Further inspiration to explore Az for optoelectronic perovskites comes from its utilization as the A-site cation in ABX$_{3}$ type ferroelectric perovskites and the consequent very distinctive properties of the single crystals attributed to Az. For example, in these perovskites large anomalies were observed in the temperature-dependent measurements of Raman modes and AC relative permittivity attributed to the ring puckering motion of an oriented azetidine during phase transitions \cite{Zhou2011_AngChem, Asaji2012_JPhysChemA, Maczka2014_InorgChem}; the latter (low frequency) was shown to reach up to $\sim$10$^{6}$ in [(CH$_{2}$)$_{3}$NH$_{2}$][Cu(HCOO)$_{3}$] \cite{Zhou2011_AngChem}. Recent work has shown the formation of a lead iodide perovskite with Az, albeit with low efficiency in a solar cell, and a mixed cation perovskite with MA which showed negligible hysteresis compared to MAPbI$_{3}$ perovskite but without any detrimental impact on the performance \cite{Pering2017_JMCA}. Moreover, as thin films and single crystals, AzPbI$_{3}$ was shown to result in much higher stability towards water compared to MAPbI$_{3}$  \cite{Pering2017_JMCA, Panetta2018_JMCA}. These initial studies suggest that Az could be a very good candidate in the search for a stable perovskite through either replacement of or partial mixing with the existing cations. \hspace*{\fill} 

\noindent
Given the ideal size of Az to form a 3D perovskite with lead halides, research into utilizing this molecule as the cation in LHPs has been, surprisingly, very limited. In this regard, we explore the utilization of Az in the synthesis of perovskite colloidal nanocrystals (NCs). NCs, owing to their facile synthesis, offer excellent handle over size control, and importantly, are intrinsically defect-tolerant \cite{Kovalenko745_Science}. Additionally, the option of color-tunability achieved through mixed halides and the corresponding bright photoluminescence spanning the visible spectrum renders NCs as excellent low-cost materials for the fabrication of LEDs and photodetectors \cite{Huang2017_ACSEnerLett}. Here, we report the synthesis of colloidal NCs using azetidine hydrochloride and lead iodide, and characterize the optoelectronic and structural properties. We determine an indirect-bandgap and an atypical structure within the NC with two distinctive interatomic distances and non-orthogonal lattice angles. Electronic transport is observed in the form of charge transfer to PCBM characterized using light-induced electron paramagnetic resonance spectroscopy (LEPR).\hspace*{\fill} 

\noindent
\textbf{Materials and Synthesis}. All procedures are carried out in air at ambient condition in a fume-hood. All chemicals were used as received without further purification. Standard hot-injection method for synthesis could not be carried out because AzCl did not dissolve in solvents used for PbI$_{2}$ even at high temperatures (upto 180\si{\degree}C) and with addition of hydroiodic acid. Hence, the synthesis was carried out by mixing and thoroughly grinding in a mortar 1 mM of lead iodide (PbI$_{2}$; 99.99$\%$; Sigma-Aldrich) and 1 mM of azetidine hydrochloride ((CH$_{2}$)$_{3}$NH$_{2}$Cl; 97$\%$; Sigma-Aldrich). Oleylamine (0.5 mL; 70$\%$; Sigma-Aldrich) and oleic acid (1 mL; 99$\%$; Sigma-Aldrich) were added as ligands to this mixture and mixed further thoroughly. The color starts changing within a minute from yellow into orange upon addition of the ligands as shown in Figure \ref{fig:Optical}(b). Further mixing was carried out to ensure complete conversion of PbI$_{2}$. Toluene (anhydrous; Sigma-Aldrich) was added to this mixture and the solution was transferred to a vial and stirred overnight. This solution, hereinafter referred to as the colloid solution (CS), was used for subsequent optoelectronic and structural characterization. Considering exact stoichiometry of the compound remains unknown, we will refer to the same in the future as "AzPbI$_{2}$Cl" perovskite assuming the structural unit is of the form ABX$_{3}$, at least based on the initial molar ratios used.\hspace*{\fill}

\begin{figure*}[t]
\includegraphics[scale=0.35]{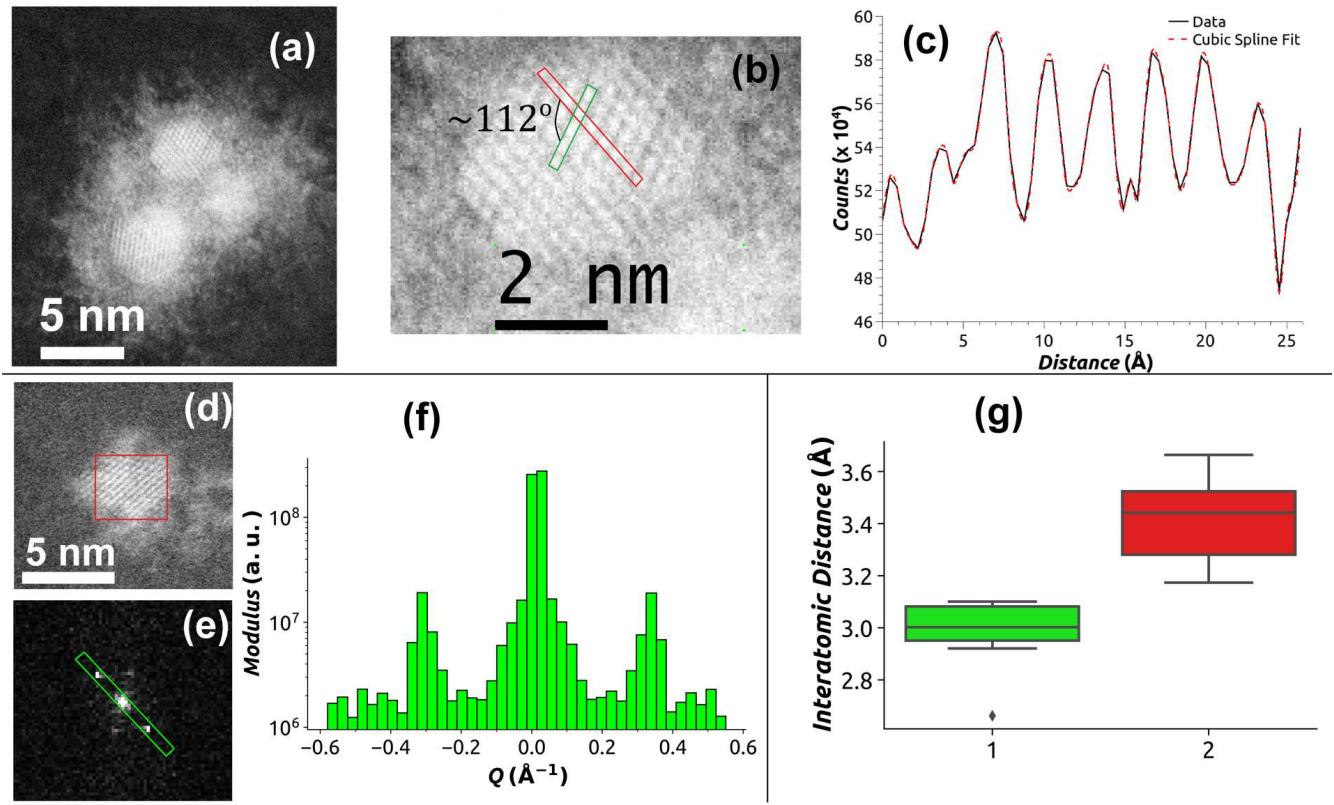}
\caption{\label{fig:TEM}(a) Transmission electron microscope (TEM) image of the nanocrystals (NCs). (b) Zoomed-in image of the top NC in (a). The red and green boxes show two examples of array of atoms whose center positions were determined by fitting cubic spline to the data as shown in plot (c) for the red box. Two distinct interatomic distances were obtained from such analysis: 2.98 $\pm$ 0.15 \si{\angstrom} and 3.43 $\pm$ 0.16  \si{\angstrom}. (d) TEM image of another NC, (e) its FFT image, and (f) bar plot of the green box in the FFT image. Analysis of the peak positions in the bar plot confirms the interatomic distances obtained in (c) as described in the text. (g) Boxplot of the interatomic distances within the nanocrystals obtained from the TEM analysis.} 
\end{figure*}

\noindent
\textbf{Structure Characterization}. Structural characterization was carried out using transmission electron microscope (TEM) imaging of the NCs. The filtrate from CS was washed twice in methyl acetate (anhydrous; Sigma-Aldrich) and drop-casted onto a carbon TEM grid and allowed to dry in ambient conditions. NC imaging and energy dispersive X-ray spectrscopy (EDS) were performed using a JEOL ARM200CF microscope operated at 200 kV accelerating voltage equipped with a 0.98Sr JEOL EDS detector. For the imaging, a $\sim$5.5 pA STEM probe with $\sim$23 mrad convergence semi-angle was used and the ADF detector collection angles spanned $\sim$50-242 mrad. Several images were recorded at a 4 $\mu$s pixel dwell time and 512x512 pixels to observe the structural change induced by the electron beam. From this image series, initial images with minimal structural change were realigned and averaged (Figure \ref{fig:TEM}(a)) using the Smart-Align software \cite{Jones2015_ASCImaging}. For the EDS, it was not possible to obtain sufficient X-ray yield from a single NC to determine the elemental composition. Therefore, a wide field of view area featuring a NC ensemble or cluster (Figure S1 in Supplementary Information) was chosen and scanned with an increased probe current of $\sim$400 pA. By increasing the probe current, significant structural damage occurred to the NCs, however, the elemental composition of the cluster was verified to contain Pb, I and Cl as discussed below.\hspace*{\fill}

\noindent
The shape of the NCs (the boundaries are blurred by the surface-coated ligands that remained after washing) appears to be spherical, but can also be partly discerned to be hexagonal. Analysis of the size of the NCs reveals the average size to be $\approx$ 4.0 nm with $\sigma\approx$ 0.3 nm (approximate because of blurred boundaries). The NCs do not appear to have a well-defined shape in contrast to the LHPs. The reason for this is unclear but could be ascribed to the kinetics of the NC formation, where incorporation of Az could possibly be the limiting factor, which is uncontrolled in our case due to the primitive synthesis method employed. \hspace*{\fill} 

\noindent
The real and reciprocal images of the NCs were analyzed to derive the structural parameters, shown in Figure \ref{fig:TEM}. In the real-space images, a row of atoms was selected, an example shown in Figure \ref{fig:TEM}(b), and the corresponding line profile (averaged over the box width) was plotted as a function of distance. This data was fitted with a cubic spline function, shown in Figure \ref{fig:TEM}(c), to determine the locations of the peak maxima which correspond to the atomic centers; cubic spline gave the best fit compared to lower order splines and Gaussian functions. With the centers determined, the interatomic distances were calculated as the difference between the centers. Peaks close to and at the border of the rows were not considered in the estimation to remove uncertainties from blurring and boundary effects. This methodology was carried out for multiple rows in two different NCs. The analysis (28 atomic centers in total from 7 rows) shows that the interatomic distances can be categorized into two distinct values (7 and 21 counts) given by mean and standard deviation: 2.98 $\pm$ 0.15 \si{\angstrom} and 3.43 $\pm$ 0.16 \si{\angstrom} respectively as shown in the boxplot of Figure \ref{fig:TEM}(g). This is further corroborated by the analysis of the Fast Fourier Transform (FFT) image of a third NC (Figures \ref{fig:TEM}(d)-(e)) with the line profile along the length of the green box shown as a bar plot in Figure \ref{fig:TEM}(f). Due to a relatively large pixel size of 0.028 \si{\angstrom}$^{-1}$ of the FFT image, only one peak at 0.31 \si{\angstrom}$^{-1}$ can be seen along the radial direction on either side of the two centers. From this, a range for the interatomic distances can be elucidated, i.e. 0.31 $\pm$ 0.028 \si{\angstrom}$^{-1}$, which corresponds to [2.96, 3.55] \si{\angstrom} - in agreement with the values obtained above from real-space analysis.\hspace*{\fill}

\noindent
Angles inherent to the NCs were derived by measuring the angle between two rows of atoms as shown with an example in Figure \ref{fig:TEM}(b). The angles are non-orthogonal and approximately 112\si{\degree} or 180\si{\degree} - 112\si{\degree} = 68\si{\degree}. The non-orthogonality of the angles can also be clearly seen in the FFT image (Figure \ref{fig:TEM}(e)), where the other radial peak is located at a non-orthogonal angle with respect to the first. From this analysis, we can see that the structure is dissimilar to that of the LHPs where cubic or orthorhombic structures are prevalent in the photoactive phases \cite{Chen2018_ChemSocRev, Rebecca2018_AcsEnerLett}.

\begin{figure*}[t]
\includegraphics[scale=0.48]{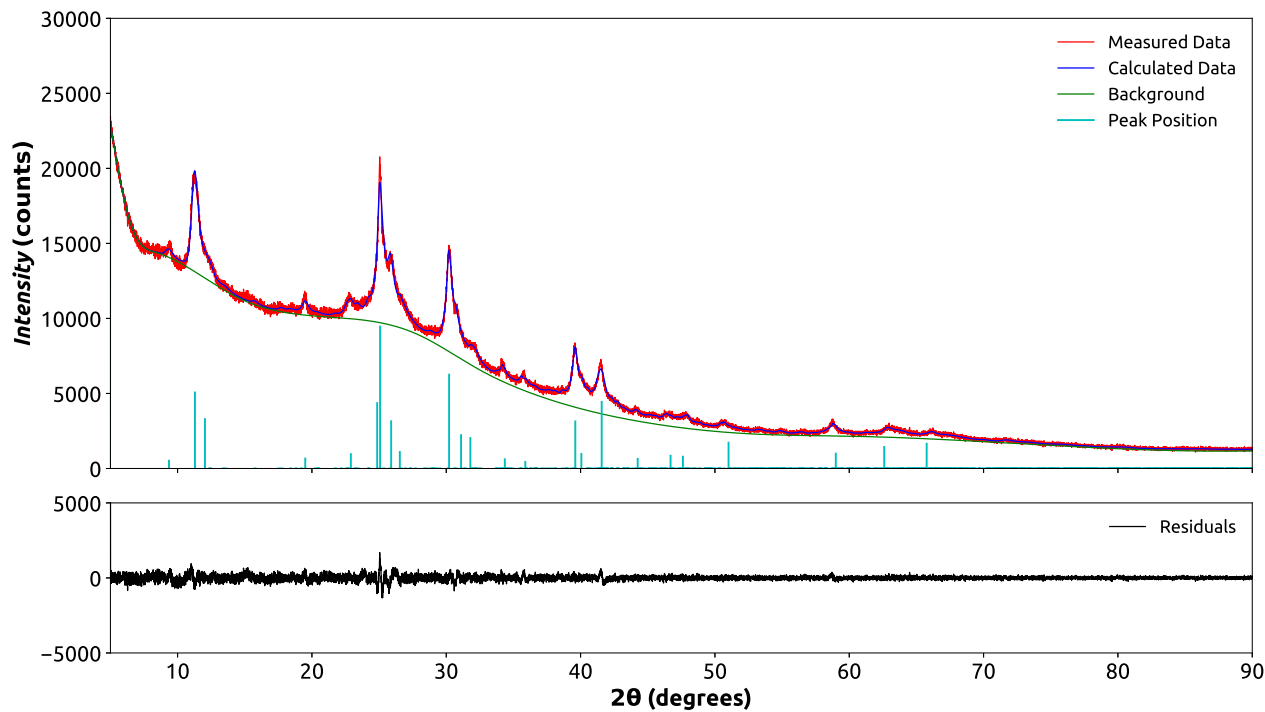}
\caption{\label{fig:XRD} XRD data and fits of the synthesized perovskite colloids. The peak positions corresponding to the diffraction planes are given in the supplementary text file.}
\end{figure*}

\noindent
X-ray diffraction (XRD) measurements were carried out to further characterize the structure. The CS was washed once in methylacetate in a centrifuge (8000 rpm for 6 min) and the precipitate was dried and collected for XRD measurements. Powder XRD data was recorded at room temperature on a Rigaku SmartLab diffractometer equipped with Cu-K$\alpha$ X-ray source with $\lambda$ = 1.54 \si{\angstrom}. Data processing, indexing and subsequent refinement were carried out in the Rigaku PDXL software. For the refinement, including the background, whole-powder-pattern-decomposition (WPPD) method was used. The data and the fits are plotted in Figure \ref{fig:XRD}. The structure obtained from the analysis is triclinic (space group P-1) with the lattice parameters $\textit{a}$ = 8.775 \si{\angstrom}, $\textit{b}$ = 10.621 \si{\angstrom}, $\textit{c}$ = 8.707 \si{\angstrom}, $\alpha$ = 116.01\si{\degree}, $\beta$ = 115.57\si{\degree}, $\gamma$ = 71.87\si{\degree}. The absence of the PbI$_{2}$ peak at 12.8\si{\degree} (peak positions in the supplementary indexing information text file) indicates its complete conversion into the product. The initial indexing search before refinement resulted in many different results, all with only one single phase, however, the above indexing gave one of the lowest fit metrics during refinement (from 5\si{\degree}-90\si{\degree}) with R$_{wp}$ = 2.29\%, R$_{p}$ = 1.70\%. The reason for choosing this indexing is that the derived lattice angles $\alpha$ and $\beta$ are in close agreement with the angle obtained from TEM analysis. Hence, we believe the derived triclinic structure to be the probable unit cell structure in our NCs. For a concrete and complete elucidation of the structure, including the type of octahedral connection, characterization of single crystals is essential but, as stated in the Synthesis section, because it was not possible to dissolve AzCl in solvents for PbI$_{2}$, single crystals could not be grown. However, structural analysis of azetidine lead halides from XRD appears to be a major challenge due to twinning and distortions reported in the case of AzPbI$_{3}$ \cite{Panetta2018_JMCA} and speculated to be between 2D and 3D \cite{Pering2017_JMCA}. To circumvent the solubility issues, alternate synthesis routes will be explored in the future. \hspace*{\fill}

\begin{figure}[t]
\includegraphics[scale=0.36]{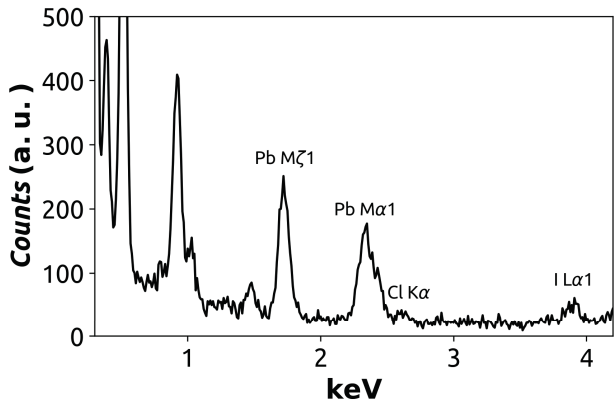}
\caption{\label{fig:EDX} EDS spectra of an ensemble of the nanocrystals (Figure S1). The spectra confirms the incorporation of chlorine into the perovskite. Estimated atomic percentages with corresponding counting statistics errors are shown in Table S1.}
\end{figure}

\noindent
EDS carried out on an ensemble of the NCs during TEM imaging verified the presence of Cl along with Pb and I (in Figure \ref{fig:EDX}). Calculated elemental compositions of the ensemble from the spectra are shown in Table S1 in atomic percent along with the \% error arising from X-ray counting statistics \cite{Watanabe2006_JofMicroscopy}. The compositions were calculated using inbuilt k-factor routines of the Thermo NSS software. We are aware that such k-factor routines could have large systematic errors upto 20\% \cite{Macarthur2016_MandM, Watanabe1996_Ultramicroscopy}. Additionally, we utilized a combination of K+L lines for the k-factor quantification in order to reduce the counting statistics error. Considering that the ionisation mechanism for each line is different, combining the lines could further add to the uncertainty in the estimation of the atomic ratios. Thus, although we were able to confirm the incorporation of Cl in the NC, the true composition, and hence the stoichiometry, remains inconclusive. To determine accurate elemental composition, a combination of EDS, EELS and known elemental standard comparison would be required \cite{Craven2016_Ultramicroscopy, Haberfehlner2014_Nanoscale, Macarthur2016_MandM, Varambhia2018_Micron, Watanabe2006_JofMicroscopy}, which will be explored in the future. \hspace*{\fill}

\begin{figure}[t]
\includegraphics[scale=0.07]{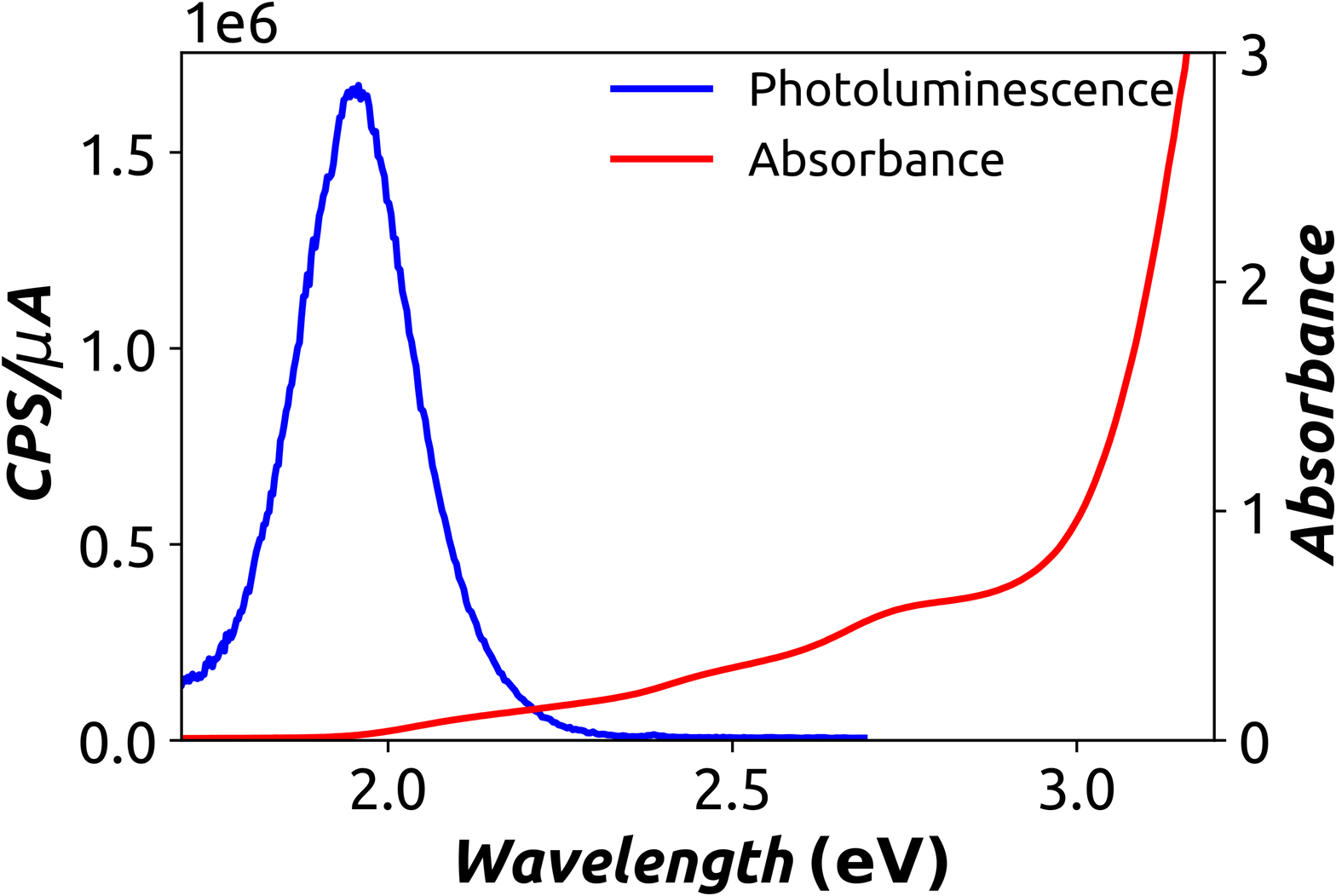}
\caption{\label{fig:PL}PL and UV-visible absorbance of the nanocrystals in solution. The band-gap is about 1.96 eV and appears of indirect nature.}
\end{figure}

\noindent
\textbf{Optical Properties}. Figure \ref{fig:Optical}(c) shows CS luminescing as bright orange color in a UV-box (365 nm). For UV-visible absorbance measurements, the CS was filtered using a 0.2 $\mu$m PTFE filter and collected in a vial (Figure \ref{fig:Optical}(c)); unlike CS which was turbid, the filtrate was a clear solution with dispersed colloids or NCs. 500 $\mu$L of the filtrate was added to a cuvette, with a path length of 1 cm, and toluene was added until the volume was \textcolor{-green!40!yellow}{3} mL. This was utilized to record steady-state photoluminescence (PL) and UV-visible absorbance spectra, shown in Figure \ref{fig:PL}. PL was measured with an automated spectrofluorometer (Fluorolog, Horiba Jobin-Yvon) equipped with a 450 W Xenon lamp excitation source and a photomultiplier tube detector; the excitation wavelength was 450 nm (2.76 eV). The absorbance measurement was recorded using a commercial Varian Cary 60 spectrophotometer.\hspace*{\fill}

\noindent
PL data in Figure \ref{fig:PL} shows typical emission with peak at $\approx$ 1.96 eV. The FWHM is about 200 meV, higher than observed with Cs/FA/MA in LHP NCs: 90-110 meV corresponding to emission peaks ranging from 1.65-3.10 eV \cite{Kovalenko745_Science, Huang2017_ACSEnerLett, Yasser2015_AdvMater}. However, since the NCs are coated with ligands, further optimization of washing steps or use of different ligands \cite{Huang2017_ACSEnerLett} could possibly help achieve a narrower bandwidth. PL data of the unfiltered CS is shown in Figure S2 and has the emission peak at the similar energy. Looking at the onset of the absorbance which converges with the PL peak energy indicating no discernible  Stokes shift, AzPbI$_{2}$Cl can be classified as an indirect-bandgap semiconductor.\hspace*{\fill}

\noindent
This is an interesting result considering that Az size is between MA and FA but forms an indirect-bandgap semiconductor with lead halides as opposed to the direct-bandgap nature of the latter \cite{Green2014_NatPhot}. The electronic band structure in LHPs is determined by the overlap of Pb and halide orbitals, which in turn is determined by the corresponding bond angles and bond distances in the unit cell, and affected by the spin-orbit coupling at the band edges \cite{Brivio2013_APLMater, Mao2018_JACS, Stoumpos2015_JACS, Azarhoosh2016_APLMat, Niesner2016_PRL, Stranks2018_NatMat, Knutson2005_InorgChem, Mao2016_ChemMater, Dohner2014_JACS}. With regards to this, it has been shown that the type of octahedron connectivity mode, i.e. corner, edge or face-sharing resulting from different dimensionality (1D, 2D or 3D), can affect the dispersity of the band structure thus determining the bandgap character, i.e. direct or indirect \cite{Mao2018_JACS, Stoumpos2017_InorgChem, Stoumpos2015_JACS, Li2015_ChemMater, Luo2018_Nature, Mao2017_JACS_2, Tran2017_MaterHor}. Moreover, First-Principle calculations have shown that the high-performing MAPbI$_{3}$ in its 3D cubic phase with corner-sharing octahedral units, has a positive enthalpy of formation and thus is thermodynamically metastable \cite{Thind2017_ChemMater}. However, its hexagonal polymorph, albeit 1D, with face-sharing octahedral units has a negative enthalpy of formation and an indirect bandgap of 2.6 eV. Thus, understanding the complete structure should provide insights into the origin of the indirect-bandgap in our material. Unfortunately, because of lack of complete structural information in our case, we could not determine the connectivity modes and so cannot definitively conclude on the origin of the indirect-bandgap in our NCs.

\noindent
The PL quantum yield turned out to be less than 0.5$\%$, and remained similar even after about 12 months after synthesis indicating stable structure; however, the NCs aggregated with time. Such low quantum yield can be attributed not only to the indirect-bandgap \cite{Mohamed2018_APL} nature but also to non-radiative recombination due to factors such as surface defects - also causing the broadening of the PL peak in Figure \ref{fig:PL}. As mentioned earlier, further optimization should improve the optical properties of the NCs. \hspace*{\fill}

\noindent
\textbf{Charge Transfer Properties}. A good electronic interaction of the NCs with other semiconducting or charge transport materials is of pivotal importance for achieving efficient organic and hybrid optoelectronic devices. An effective way to probe whether NCs can “communicate” efficiently with other molecules is the investigation of their electron transfer capabilities. To get a direct view of the electron transfer process, we applied light-induced EPR spectroscopy (LEPR). LEPR spectroscopy is a well-suited technique to detect long-living paramagnetic species (such as radical anions and cations) generated after visible light absorption and electron transfer processes \cite{Niklas2017_AENM, Krause2014_SolEnerMater, Krinichnyi2011_AIPAdv}. \hspace*{\fill}

\noindent
The LEPR spectra were recorded at 80 K and 280 K on a Bruker Elexsys E580 X-band spectrometer equipped with a nitrogen gas-flow cryostat for sample temperature control. The sample temperature was maintained with an Oxford Instruments CF9350 cryostat and controlled with an Oxford Instruments ITC503. LEPR spectra were acquired using white light illumination from a xenon lamp, IR filtered and focused onto a quartz optical fibre ending in the optical window of the EPR spectrometer cavity. The system delivered about 50 mW cm$^{-2}$ of light irradiance to the sample. The LEPR experimental parameters were a modulation amplitude of 1 G and a microwave power of 0.2 mW. LEPR measurements were performed on thick films of neat PCBM (as control) and PCBM:NCs blend. As for the neat PCBM (99\%; Solenne B.V.), a 2 mg mL$^{-1}$ solution of PCBM in toluene was prepared. As for the PCBM:NCs blend, a PCBM solution in toluene (8 mg mL$^{-1}$) was mixed with an equal volume of NCs solution (from filtered CS) in toluene. The final solutions were poured inside a quartz EPR tube (inner diameter of 3 mm) and evaporated under vacuum, leaving a  thick film on the inner tube wall. Subsequently, EPR tubes were sealed under vacuum. We chose PCBM since it is the most common n-type material used in perovskites and often used as an electron-acceptor in many optoelectronic applications \cite{Ostroverkhova2016_ChemRev}. The LEPR measurements are performed at 80 K to slow down the recombination dynamics of the photogenerated charges and have a stronger EPR signal.\hspace*{\fill} 

\begin{figure}[t]
\includegraphics[scale=0.65]{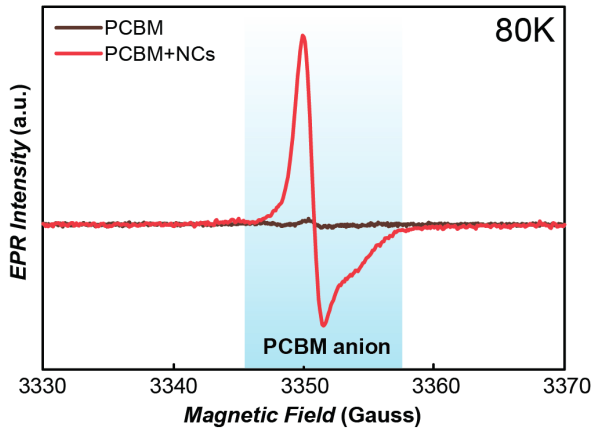}
\caption{\label{fig:EPR}LEPR spectra of PCBM (brown) and PCBM:NCs blend (red) films acquired at 80 K. The spectra shown are the difference between spectra acquired under visible light (light ON) and before (dark) sample illumination.}
\end{figure}

\noindent
The LEPR spectra (light ON minus dark) recorded at 80 K in the neat PCBM film and the PCBM:NCs blend are reported in Figure \ref{fig:EPR}. Before light illumination (dark) no EPR signal is detected in any sample underlining that no paramagnetic species are present in the ground state. After visible light irradiation of the sample, a sharp EPR signal is detected in PCBM:NCs blend. The signal is characterized by a g-factor (g = 1.9998 $\pm$ 0.0005), a linewidth ($\Delta$B$_{pp}$ $\approx$ 1 G) and an anisotropic lineshape typical of the long-living radical anion localized on PCBM \cite{Righetto2018_Nanoscale}. The generation of the PCBM radical anion after visible light illumination of the PCBM/perovskite NCs blend was already revealed in literature in the case of MAPbI$_{3}$ and it is usually attributed to the photoinduced electron transfer process \cite{Righetto2018_Nanoscale, Privitera2017_MRSAdv}. The same mechanism can take place in our PCBM:NCs blend. After light absorption, an electron transfer between NCs (electron-donor) and the PCBM (electron-acceptor) occurs leaving a radical anion localized on PCBM and a hole localized on the NC. Bearing that S=1/2, the hole should be detectable through EPR spectroscopy. Nevertheless, its EPR spectrum is not detected probably because of a severe line broadening due to fast magnetic relaxation times favoured by the spin-orbit coupling interaction of the hole with the heavy nuclei of our NCs \cite{Privitera2017_MRSAdv}. Conversely to that observed in our PCBM:NCs blend, the neat PCBM film does not show any photoinduced signal corroborating that the electron transfer does not occur between PCBM molecules but only between NCs and PCBM. The same EPR analysis was also performed at 280 K to rule out the effect of temperature and phase transitions on the electron transfer process. The spectrum reported in Figure S3 shows the same results of the 80 K analysis. The lower EPR intensity at 280 K can be rationalized considering different factors. First, since LEPR spectroscopy is a steady-state technique, the signal intensity is given by a balance between generation and recombination of the charge carriers. At higher temperature, charge recombination is faster and therefore the LEPR signal is lower \cite{Niklas2017_AENM}. Second, the EPR signal is proportional to the paramagnetic susceptibility which follows the Curie law and so is inversely proportional to the temperature \cite{solidstate}. Thus, at higher temperatures, the intensity is lower. Finally, spin relaxation times also influence EPR intensity and being faster at higher temperatures, result in a decrease in the signal intensity \cite{Niklas2017_AENM}. \hspace*{\fill}

\noindent
The LEPR analysis provided a direct confirmation that a strong electronic interaction, resulting in charge transfer and charge separation, occurs between NCs and PCBM suggesting potential of our NCs in applications relevant for optoelectronics. From these observations and the high low frequency dielectric constants imparted by Az to the ferroelectric perovskites \cite{Zhou2011_AngChem}, we expect charge generation to be efficient. The high dielectric constants in the ferroelectric perovskites has been attributed to the polarization caused by a highly oriented Az \cite{Zhou2011_AngChem, Asaji2012_JPhysChemA, Maczka2014_InorgChem} in contrast to the dynamic orientational disorder of linear molecules MA and FA \cite{Hutter2017_NatMater, Motta2015_NatComm, Fan2015_JPCL}. Theoretical calculations on MAPbI$_{3}$ show low barrier for the rotation of MA cation leading to dynamic orientational disorder and thus no long-range ordering of the dipoles for ferroelectric effect under electric field \cite{Brivio2013_APLMater, Fan2015_JPCL}. While MA and FA are linear molecules possessing 3D degrees of rotational freedom \cite{Hutter2017_NatMater, Motta2015_NatComm, Fan2015_JPCL}, azetidine, however, is a four-membered ring with bond angles much more constricted compared to higher order rings (Figure \ref{fig:Optical}(a)). The ring structure restricts torsional motion of the atoms thereby reducing the rotational degrees of freedom. When constrained in a cage-like structure as in a perovskite and ionically bound to the halides, this is even more the case. Consequently, the abnormally high dielectric constants due to high polarization reported in \cite{Zhou2011_AngChem} and attributed to the ring-puckering of Az can only occur if Az is highly oriented in the perovskite structure. Hence, presuming the case to be similar in our NCs, this constraint should induce a relatively highly oriented dipole compared to MA and FA and thus, plausibly lead to efficient charge transfer as a result of a strong screening effect.

\noindent
\textbf{Conclusions}. Our work has shown the suitability of azetidine in forming perovskite-type NCs with lead iodide and revealing properties with potential for optoelectronic applications. Interestingly, the synthesized NCs result from the chlorine-adduct of azetidine rather than the iodine or bromine. TEM analysis of the NCs showed two distinctive interatomic distances, 2.98 \si{\angstrom} and 3.43 \si{\angstrom} and lattice angle of $\approx$ 112\si{\degree}. XRD characterization revealed a probable triclinic unit cell with the lattice angles in agreement with the TEM analysis but needs confirmation through characterization of a single crystal. LEPR measurements verified the electron transfer capabilities of the NCs through photoinduced charge transfer to PCBM. While EDS of the NCs confirmed the presence of chlorine, accurate composition could not be verified. Nevertheless, this points towards the formation of lead mixed halide perovskite with chlorine and iodine observed for the first time in NCs, and not seen before with Cs, MA or FA cations, with azetidinium as the cation which seems to play a role in driving such formation, implying that not just the size (determining the tolerance factor) but also the geometry of the molecule, affecting the rotational degrees of freedom, could play a role in perovskite formation. The indirect-bandgap of 1.96 eV displayed in AzPbI$_{2}$Cl unlike the direct-bandgap nature of Cs/MA/FA lead halide perovskites, may entail the need for thick films for applications in solar cells. However, mixed cation combinations with Cs/MA/FA can be explored \cite{Pering2017_JMCA} to develop systems with excellent photovoltaic performance (owing to Cs/MA/FA) and high stability (negligible hysteresis and water resistant owing to azetidine) \cite{Pering2017_JMCA, Panetta2018_JMCA}. Along with the latter, systematic studies, including developing azeditine analogs with inorganic elements, open doors to understand and subsequently harness the high AC dielectric constants endowed by azetidine (attributed to its ring-puckering motion \cite{Asaji2012_JPhysChemA, Zhou2011_AngChem, Maczka2014_InorgChem} due to its geometry) to the ferroelectric perovskites in lead halide perovskites potentially leading to a paradigm-shift in the pursuit of stable, high efficiency perovskite optoelectronic technologies.\hspace*{\fill}

\noindent
\textbf{Supplementary Information}. TEM image of NC cluster; elemental compositions calculated from EDS spectra of the NCs; PL spectra of unfiltered colloid solution and LEPR spectra of the NCs measured at 280 K.\hspace*{\fill}

\noindent
\textbf{Indexing Information}. The text file contains the peak positions and the corresponding intensities obtaining from the XRD data analysis.\hspace*{\fill}

\noindent
\textbf{Acknowledgements}. SVK would like to thank Dr. Amir-Abbas Haghighirad and Dr. Dharmalingam Prabharakaran for assistance with alternative synthesis routes and corresponding XRD. We thank Centre for Advanced Electron Spin Resonance (CAESR), Department of Chemistry, University of Oxford for EPR measurements. The work at CAESR was supported by the EPSRC (EP/L011972/1). AP also thanks Dr. William Myers, CAESR facility for his kind assistance with EPR measurements. TEM/EDX work was performed on the South of England Electron Microscope funded through EPSRC (grant EP/K040375/1). SVK expresses gratitude to EPSRC (WAFT, grant EP/M015173/1) and UKRI (START, grant ST/R002754/1) and AP to European Union's Horizon 2020 research and innovation programme (SEPOMO, Marie Sklodowska Curie grant agreement no. 722651) for funding.\hspace*{\fill}

\noindent
\textbf{Author Contributions}. SVK conceived and directed the study with MKR and HJS. YH worked with SVK on synthesis of NCs, optical and XRD measurements, and prepared the TEM samples. AP measured and analyzed the EPR data, and prepared the EPR part of the manuscript. AV carried out the TEM imaging and EDS spectroscopy; and analysed the EDS spectra. Analyses of the optical, TEM and XRD data was carried out by SVK. The manuscript was mainly prepared by SVK and all authors participated in the preparation and review.\hspace*{\fill}

\noindent
\textbf{Competing Interests}. The authors declare the following competing financial interest: HJS is a founding member of Oxford Photovoltaics ltd and Heliochrome ltd, which seeks to commericalize perovskite solar cells and LEDs.\hspace*{\fill}

\bibliographystyle{apsrev}

\end{document}